\documentclass{ws-p8-50x6-00}
\def\PRB{{\em Phys. Rev.} B}
\def\NIMA{{\em Nucl.~Instrum.~Methods} A}

\newcommand{\GeVcc}    {\mbox{$ {\mathrm{GeV}}/c^2                           $}}
\newcommand{\GeVccv}    {\mbox{$ {\mathrm{GeV}}/c^2                           $}~}

\newcommand{\hetrois}    {\mbox{$ ^{3}{\mathrm{He}}                            $}~}
\newcommand{\hetro}    {\mbox{$ ^{3}{\mathrm{He}}                            $}}

\newcommand{\neut}{$\tilde{\chi}$~}
\newcommand{\neutt}{$\tilde{\chi}$}

\newcommand{\gam} {{$\gamma$-ray}~} 
\newcommand{\gams} {{$\gamma$-rays}~} 
\newcommand{\gamss} {{$\gamma$-rays}}
\begin{document}

\title{A superfluid \hetrois detector for direct dark matter search}

\author{F. Mayet $^{1}$, D. Santos $^{1}$, 
Yu. M. Bunkov $^{2}$, G. Duhamel $^{1}$,}
\author{H. Godfrin $^{2}$, F. Naraghi $^{1}$, G. Perrin $^{1}$}
%
%
\address{$^{1}$ Institut des Sciences Nucl\'eaires, 
CNRS/IN2P3 and Universit\'e Joseph Fourier,\\
 53, avenue des Martyrs, 38026 Grenoble cedex, France}
\address{$^{2}$ Centre de Recherches sur les Tr\`es Basses Temp\'eratures, \\
 CNRS, BP166, 38042 Grenoble cedex 9, France} 


\maketitle

\abstracts{MACHe3 (MAtrix of Cells of superfluid \hetro) is a project of a new detector for 
direct Dark Matter Search. The idea is to use superfluid \hetrois as a sensitive medium. The existing device, the
superfluid \hetrois cell, will be briefly introduced. Then a description of the MACHe3 project will be presented, in
particular the background rejection and the neutralino event rate that may be achieved with such a device.}

\section{Introduction}
\begin{figure}[hbt]
\begin{center}
\epsfxsize=13.5pc 
\epsfbox{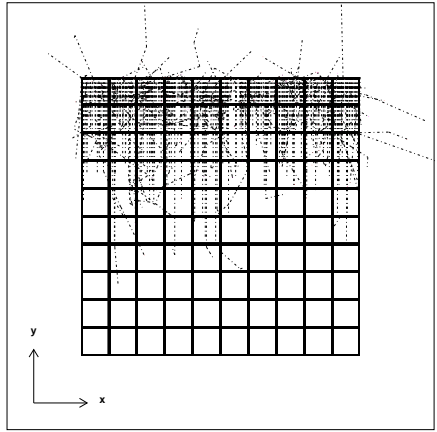}
\caption[]{2-dimensional view of a proposed matrix of 1000 cells of \hetrois (125 $cm^{3}$ each). 
The events,  generated in a direction perpendicular to the upper face, simulate 10 keV neutrons.}
\label{fig:mat}
\end{center}
\end{figure}

Many existing experiments for direct Dark Matter (DM) search \footnote{In particular, 
we shall suppose all through this work a neutralino (\neutt), the lightest supersymmetric particle,
 as the particle making up the bulk of galactic cold DM.} are imposing upper limits on the interaction cross-section 
of neutralino (\neutt) on proton, both scalar and axial. They use a great variety of materials and detection techniques. 
However, the neutrons are still very difficult to 
discriminate along with a good rejection for \gamss. The cosmogonic activation is, in all these experiments, an important
source of intrinsic contamination. 
There are several advantages of using \hetrois as a sensitive medium for direct DM detection. Some of them are :
\begin{itemize}
\item It presents a high neutron capture cross-section (up to $\sim \!10^3$ barn for $\sim \!0.1$ eV thermal neutrons), giving a clear neutron 
signature and hence a good rejection, as shown in sec \ref{sec:back}. As neutrons interact {\it a priori} like \neutt, this gives to 
an \hetrois detector the possibility to discriminate the \neut signal from its ultimate background noise.
\item Superfluid \hetrois is produced with a very high purity, making this medium almost free of radioactive
contaminations.
\item The interaction of a \neut on the \hetrois nucleus (odd spin) is mainly a spin-dependent one, 
making this medium complementary to other media used in existing DM experiments.
\end{itemize}
In addition a DM detector should present an energy threshold as low as possible.  
At ultra low temperatures (T $\simeq \!100 \,\mu\mathrm{K}$), \hetrois in the superfluid phase B presents in a
Lancaster's cell configuration \cite{lanc} a low detection threshold ($\mathrm{E}_{th}\simeq 1 \,\mathrm{keV}$).\\  
In order to enhance the rejection against the background (neutrons, \gams and muons), a high granularity superfluid
\hetrois detector (fig.~\ref{fig:mat}) as been proposed~\cite{firstmac3,dm20}. This matrix configuration would allow to obtain a high 
background rejection (sec~\ref{sec:back}) along with a reasonable \neut event rate (sec.~\ref{sec:susy}).

\section{The superfluid \hetrois cell}
The primary device \cite{prl95} consists of a small copper cubic box ($\mathrm{V}\simeq 0.125 \;\mathrm{cm}^{3}$)
 filled with superfluid \hetro. It is immersed in a larger
volume containing liquid \hetrois and thin plates of copper nuclear-cooling 
refrigerant. Two vibrating wires are placed inside the cell, forming a Lancaster type bolometer \cite{prl95}. 
The detection principle is the following : the energy released, in the cell by the incoming particle, creates a 
cloud of \hetrois quasiparticles. 
This cloud produces a damping effect on the vibrating wire, allowing the energy released in the cell to be measured.
A small hole on one of the box walls connects the box to the main \hetrois volume, allowing the diffusion 
of the thermal excitations.\\
Although the primary experiment \cite{prl95} was still rudimentary, it has allowed to detect 
signals down to 1 keV.
\section{MACHe3 :  MAtrix of Cells of superfluid \hetrois}
The performance of a detector for direct DM search is closely related to its \neut event rate and its rejection power
against background events (thermal and fast neutrons, muons and \gamss). 
This section describes a \neut rate estimation (sec.~\ref{sec:susy}) as well as a
Monte Carlo background simulation (sec.~\ref{sec:back}).
\subsection{\neut rate estimation}
\label{sec:susy}
\begin{figure}[ht]
\begin{center}
\epsfxsize=22pc 
\epsfbox{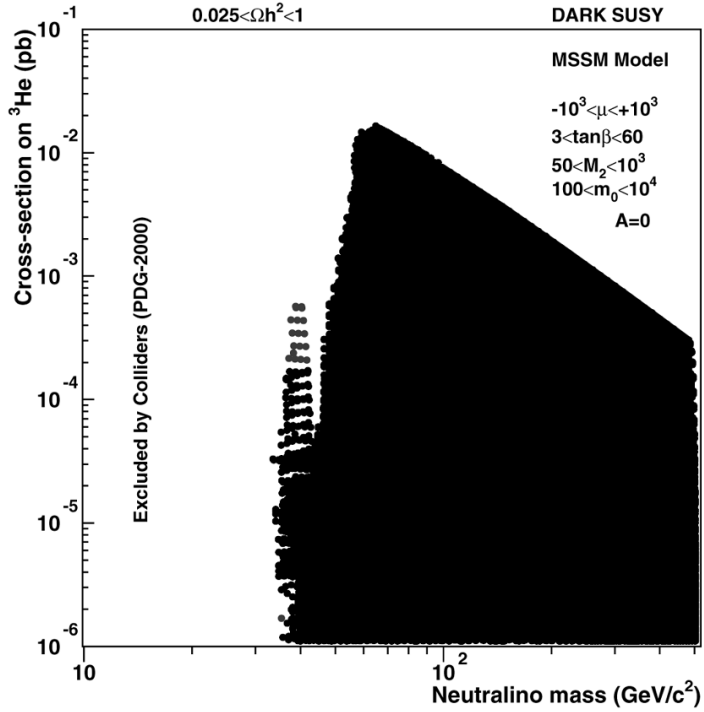}
\caption{Cross-section on \hetrois (pb) as a function of the \neut mass (\GeVcc). All points have been checked to give a \neut cosmological
relic density in the range of interest and not to be excluded by collider experiments.
\label{fig:susy}}
\end{center}
\end{figure}

At the tree-level, the spin-dependent elastic scaterring of \neut on quark is done via a squark or Z$^0$ exchange. 
The amplitude on proton (neutron) is calculated by adding the contribution of each quark, 
weighted by the quark contents of the nucleon. The axial cross-section on \hetrois is
given by :
\begin{equation}
\sigma_{spin}(^{3}{\mathrm{He}}) = \frac{32}{\pi} G^2_F m_r^2 \frac{(J+1)}{J} \left(a_p \!<\!S_p\!> + \;a_n  \!<\!S_n\!> \right)^2
\label{eq:xs} 
\end{equation}
where $a_{p/n}$ is the amplitude on proton (neutron), $<\!S_{p/n}\!>$ the spin contents 
of the \hetrois nucleus ($<\!S_p\!> = \!-0.05$ and $<\!S_n\!> = \!0.49$), $m_r$ is the reduced mass and $J$ the \hetrois spin.\\
A large scan of the supersymmetric (SUSY) parameter space has been done using the DARK SUSY code~\cite{ds}, in a constrained MSSM model. Each point have been checked to satisfy 
the following conditions : 
{\bf i)} not to be excluded by the latest collider constraints 
{\bf ii)} to give a \neut relic density in the range : $0.025 \leq \Omega_{\chi} h^2 \leq 1$.  
Fig.~\ref{fig:susy} presents the cross-section on \hetrois as a function of the \neut mass, for all points satisfying the two conditions.
The event rate is evaluated \footnote{It has been checked that, for the \hetrois nucleus, the scalar cross-section is always negligible 
compared to the axial cross-section, at least in the MSSM parameter space giving a reasonable 
event rate. Consequently, the value used for R is : $\sigma_{tot} = \sigma_{spin}$}
 as follows~:
\begin{equation}
\rm{R}\!=\! \frac{\sigma_{tot}(^{3}{\mathrm{He}})}{M_{\chi}}  \!\times \rho_{halo}\!\times\! v_0 \!
\times\! \frac{M_{det}}{M_{He}}
\label{eq:rate} 
\end{equation}
with  $\rho_{halo} = 0.3 \,\mathrm{GeV}c^{-2} \mathrm{cm}^{-3}$ and $v_0 = 270 \,\mathrm{km}.\mathrm{s}^{-1}$. 
The mass of the detector is M$_{det} = 10 \,\mathrm{kg}$, as obtained by the optimization of the background rejection (see sec.~\ref{sec:back}).\\
This leads to a maximum rate $\mathrm{R} \simeq 0.5 \,\mathrm{day}^{-1}$. 
A rate higher than $\sim 0.1\, \mathrm{day}^{-1}$ is
achieved for a large number of SUSY models with a mass of \neut in the 50-100 \GeVccv range.\\
In addition, a \neut event in MACHe3 would release a maximum energy of :
\begin{equation}
\mathrm{E}_{recoil}^{max} = 2\! \times\! \frac{\rm{M}_{\rm{He}} \rm{M}_{\chi}^2}{\left(\rm{M_{He}  \!+\!  M_{\chi}}\right)^2}
\!\times\!  v_0^2 \simeq  2 \mathrm{M_{He}}  v_0^2 \simeq 6\; \mathrm{keV}
\label{eq:max}
\end{equation}
Consequently a \neut event in MACHe3 will have the following characteristics: only one cell fired (single-cell 
event), due to the low cross-section ($\simeq 10^{-2}$~pb), and less than 6 keV in this cell.

\subsection{Background rejection simulation}
\label{sec:back}
In order to estimate the background rejection, one has to evaluate the number of background events giving 
a \neut false event (less than 6 keV in one cell). For this purpose a complete Monte-Carlo simulation has been done, 
using GEANT3.21 package and in particular the GCALOR-MICAP(1.04/10) package for slow neutrons.
The simulated detector (fig.~\ref{fig:mat}) consists of a cube containing a variable number of cubic \hetrois cells, immersed 
in a large volume containing \hetro. 
Each cell is surrounded by a thin copper layer (0.1 $mm$) and separated from the others by a 2 $mm$ gap (filled with \hetro).\\
It has been shown \cite{firstmac3} that the best design for MACHe3, as far as background rejection 
is considered, would be a matrix of 1000 cells of 5 cm side each.\\
The rejection against background events is achieved by choosing only events having the following characteristics : 
{\bf i)} Only one cell fired (single-cell event) 
{\bf ii)} Energy measurement in the cell below 6 keV 
{\bf iii)} The fired cell in the inner part of the matrix (veto selection), 
thus improving the rejection for low energy neutrons interacting elastically.
\begin{figure}[t]
\begin{center}
\epsfxsize=20pc 
\epsfbox{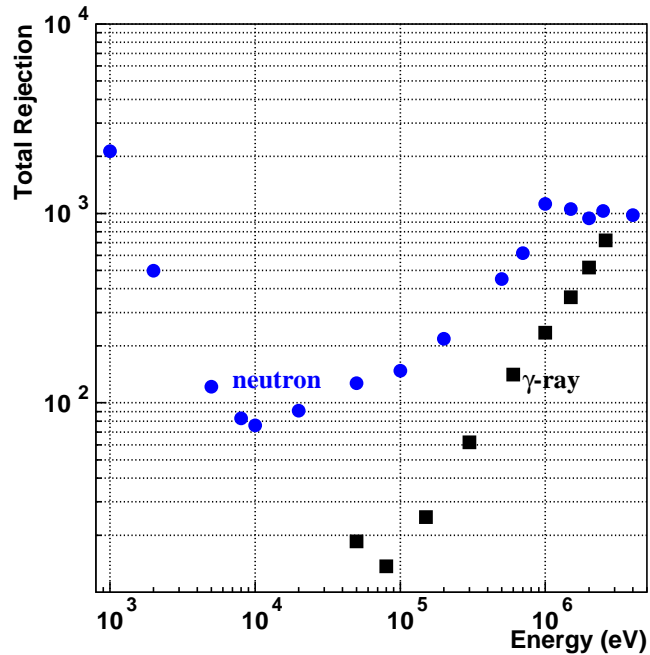} 
\caption{Background Rejection as a function of the incident particle energy, neutrons (circles) and \gams (squares), for a matrix of 1000 
cells (125
$cm^{3}$ each).
\label{fig:rejec}}
\end{center}
\end{figure}
For the preferred design the background rejection \footnote{ The background rejection is defined as the ratio 
between the number of incoming particles and the number of false \neut events 
(less than 6 keV in one non-peripheric cell).} has been obtained for neutrons and \gams for several values of kinetic 
energies (fig.~\ref{fig:rejec}).\\
For \gamss, the rejection ranges between 20 and 800, depending on the \gam energy. This indicates that a good intrinsic rejection can be
obtained with the proposed matrix configuration.\\
At low energy, the neutron rejection is dominated by the capture process, whereas high energy neutrons (1 MeV) are mainly
rejected due to the fact that the probability to leave less than 6 keV in the cell is decreasing with increasing neutron energy.\\
Using the measured neutron flux in the Laboratoire Souterrain de Modane (LSM)\cite{Chazal:1998qn} in the 2-6 MeV range 
($\Phi_{n}\simeq 4\times 10^{-6} \,cm^{-2}s^{-1}$), and the simulated rejection, it is possible to evaluate the false event rate
induced by neutrons~\cite{firstmac3}. For this purpose a simulation of a 30 $cm$ wide paraffin shielding has also been done. 
We found \cite{firstmac3} an overall neutron flux through the shielding of 5.1$\times 10^{-8} \,cm^{-2}s^{-1}$, 
with the neutron kinetic energy ranging between $10^{-2}$~eV and 6 MeV. Using this flux and the expected rejection factor 
(fig.~\ref{fig:rejec}), the false event rate induced by neutron background is estimated to $\sim\, 0.1\, \mathrm{day}^{-1}$ through the
10 kg matrix (1000 cells of $125 \,cm^{3}$). The same simulation has been done for muons (200 GeV). Most of them are 
interacting in all crossed cells. We found \cite{firstmac3} a false event rate of the order of $0.014 \,\mathrm{day}^{-1}$ in the proposed 
matrix configuration.

\section{Conclusion}
MACHe3 would allow to obtain a high intrinsic rejection against $\gamma$-ray, neutron and muon background, 
by means of correlation among the cells and energy loss measurement.  
The contamination is evaluated, for $\mu$ and neutrons, as well below the maximum expected \neut rate. 
The results of the simulation are encouraging, both for signal and background rejection.

\end{document}